\documentclass[conference, 9pt]{IEEEtran}

\usepackage{graphicx}
\usepackage{color}
\usepackage{placeins}
\usepackage{float}
\usepackage{tabularx,colortbl}
\usepackage{amssymb}
\usepackage{amsthm}
\usepackage{cite}
\usepackage{amsmath}
\usepackage{caption}
\usepackage{subcaption}

\def\vectorize{\mathrm{vec}}

\def\kron{\otimes}
\def\tr{\mathrm{tr}}

\def\vectorize{\mathrm{vec}}

\def\Htran{\mbox{\tiny H}}

\newcommand{\fracSum}[1]{{\underset{{#1}}{\sum}}}

\newcommand{\fracSumtwo}[2]{\overset{#2}{\underset{#1}{\sum}}}
\newcommand{\vect}[1]{\mathbf{#1}}

\theoremstyle{remark}

\newtheorem{theorem}{Theorem}
\newtheorem{corollary}{Corollary}
\newtheorem{lemma}{Lemma}

\begin{document}

\title{Optimizing Multi-Cell Massive MIMO for Spectral Efficiency: \\ How Many Users Should Be Scheduled?}

\IEEEoverridecommandlockouts

\author{\IEEEauthorblockN{Emil~Bj\"ornson\IEEEauthorrefmark{1}\IEEEauthorrefmark{2}\IEEEauthorrefmark{3},
Erik G.~Larsson\IEEEauthorrefmark{1}, and
M\'erouane~Debbah\IEEEauthorrefmark{3}}
\IEEEauthorblockA{\IEEEauthorrefmark{1}Department of Electrical Engineering (ISY), Link\"{o}ping University, Link\"{o}ping, Sweden}
\IEEEauthorblockA{\IEEEauthorrefmark{2}SUPELEC, Gif-sur-Yvette, France, and \IEEEauthorrefmark{3}ACCESS, Dept.~of Signal Processing, KTH, Stockholm, Sweden}%
\thanks{This research has received funding from the EU FP7 under ICT-619086 (MAMMOET), Swedish Research Council, and ERC Grant 305123 MORE.}}

\maketitle

\begin{abstract}
Massive MIMO is a promising technique to increase the spectral efficiency of cellular networks, by deploying antenna arrays with hundreds or thousands of active elements at the base stations and performing coherent beamforming. A common rule-of-thumb is that these systems should have an order of magnitude more antennas, $N$, than scheduled users, $K$, because the users' channels are then likely to be quasi-orthogonal. However, it has not been proved that this rule-of-thumb actually maximizes the spectral efficiency. In this paper, we analyze how the optimal number of scheduled users, $K^\star$, depends on $N$ and other system parameters. The value of $K^\star$ in the large-$N$ regime is derived in closed form, while simulations are used to show what happens at finite $N$, in different interference scenarios, and for different beamforming.
\end{abstract}


\IEEEpeerreviewmaketitle

\vspace{-0.1cm}

\section{Introduction}

Cellular communication networks are continuously evolving to keep up with the rapidly increasing demand for wireless data services. Higher area throughput (in bit/s per unit area) has traditionally been achieved by a combination of three multiplicative factors \cite{Nokia2011}: more frequency spectrum (Hz), higher cell density (more cells per unit area), and higher spectral efficiency (bit/s/Hz/cell). This paper considers the latter and especially the massive multiple-input multiple-output (MIMO) concept, proposed in \cite{Marzetta2010a}, which is identified as a key to increase spectral efficiency (SE) by orders of magnitude \cite{Baldemair2013a,Boccardi2014a,Larsson2014a}.

The massive MIMO concept is based on equipping base stations (BSs) with hundreds or thousands of antenna elements which, unlike conventional cellular technology, are operated in a coherent fashion. This can provide unprecedented array gains and a spatial resolution that allows for multi-user MIMO communication to tens or hundreds of user equipments (UEs) per cell, while maintaining robustness to inter-user interference. The research on massive MIMO has so far focused on establishing the fundamental physical layer properties; in particular, that the acquisition of channel state information (CSI) is limited by the channel coherence and how this impacts SEs and the ability to mitigate inter-cell interference \cite{Marzetta2010a,Jose2011b,Hoydis2013a}. The improvements in energy efficiency were analyzed in \cite{Ngo2013a,Bjornson2015a}, while \cite{Bjornson2014a,Bjornson2015b} predict that the hardware impairments of practical transceivers has small impact on the SE. In contrast, the research community has only briefly touched on resource allocation problems in the medium access control layer (e.g., user scheduling)---although the truly achievable SE can only be understood if the layers are jointly optimized.

The importance of resource allocation for massive MIMO was described in \cite{Huh2012a}, where initial guidelines were given. A main insight is that the limited number of orthogonal pilot sequences need to be allocated among the UEs to limit interference---by capitalizing on pathloss differences \cite{Li2012a,Mueller2013a} and spatial correlation \cite{Yin2013a,Li2013a}.

In this paper, we consider a related resource allocation question: \emph{how many UEs should be scheduled per cell to maximize the spectral efficiency?} This question has, to the best of our knowledge, not been answered for multi-cell systems.\footnote{A few results for single-cell systems are available, for example, in \cite{Ngo2013a}.} We show how the coherence block length, number of antennas, and other system parameters determine the answer. To this end, we derive new SE expressions for the uplink with power control and for two different beamforming schemes.

\section{System Model}

This paper considers a cellular network with universal time and frequency reuse. Each cell is given an index in the set $\mathcal{B}$, where the cardinality $|\mathcal{B}|$ is the number of cells. The BS in each cell is equipped with an array of $N$ antennas and communicates in the uplink with $K$ single-antenna UEs in the cell. We consider a massive MIMO topology where $N$ is large and fixed, while $K$ is chosen adaptively.

Each UE is picked at random from an arbitrary user distribution in the serving cell. The geometric position of UE $k \in \{1,\ldots,K\}$ in cell $l\in \mathcal{B}$ is thus a random variable and denoted by $\vect{z}_{lk} \in \mathbb{R}^2$. The channels are modeled as block-fading; $\vect{h}_{jlk} \in \mathbb{C}^N$ denotes the channel response between the considered UE and BS $j$ in a given coherence block of duration $T$ channel uses. This channel varies independently between one coherence block and the next. The realizations are drawn from circularly symmetric complex Gaussian distributions, \vspace{-2mm}
\begin{equation} \label{eq:channel-distribution}
\vect{h}_{jlk} \sim \mathcal{CN}\Big(\vect{0},d_j(\vect{z}_{lk}) \vect{I}_N \Big),
\end{equation}
which is a reasonable model for arrays with both few and many antennas (see recent measurements reported in \cite{Gao2015a}). The deterministic function $d_j(\vect{z})$ gives the variance of the channel attenuation from BS $j$ at any arbitrary UE position $\vect{z}$. We assume that $d_j(\vect{z}_{lk})$ is known at BS $j$ for all $l$ and $k$, while the exact UE positions are unknown.

The received signal $\vect{Y}_j \in \mathbb{C}^{N \times T}$ at BS $j \in \mathcal{B}$ in a coherence block is modeled, similarly to \cite{Marzetta2010a,Hoydis2013a,Ngo2013a}, as \vspace{-1mm}
\begin{equation} \label{eq:system-model}
\vect{Y}_j = \sum_{l \in \mathcal{B}} \sum_{k=1}^{K} \sqrt{p_{lk}} \vect{h}_{jlk} \vect{x}_{lk}^{\Htran} + \vect{N}_{j}
\end{equation}
where $\vect{x}_{lk}^{\Htran} \in \mathbb{C}^{1 \times T}$ is the vector signal from UE $k$ in cell $l$ (normalized such that $\mathbb{E}\{ | [\vect{x}_{lk}]_t |^2 \} = 1$ for all elements $t \in \{1,\ldots,T\}$), $p_{lk}$ is the corresponding transmit power, and $\vect{N}_{j}  \in \mathbb{C}^{N \times T}$ is additive noise with each element distributed independently as $\mathcal{CN}(0,\sigma^2)$.

Contrary to most previous works on massive MIMO, which assume fixed uplink power, we consider statistics-aware power control\footnote{Channel-aware power control was considered in \cite{Guo2014a,Bjornson2015a}, but requires a feedback mechanism for instantaneous CSI. Since small-scale fading average out in massive MIMO, we expect statistical power control to be sufficient.}; the signals from UE $k$ in cell $l$ are allocated the power $p_{lk} = \frac{\rho}{d_l(\vect{z}_{lk})}$, where $\rho>0$ is a design parameter. This power-control policy inverts the average channel attenuation and has the merit of making the average effective channel gain the same for all UEs: $\mathbb{E}\{ p_{lk} \|\vect{h}_{llk}\|^2 \} = N \rho$. This policy guarantees a uniform user experience, saves valuable energy at UEs, and avoids near-far blockage where the receiver's limited dynamic range makes weak signals drown in stronger signals.

\section{Average Per-Cell Spectral Efficiency}

In this section, we derive and analyze the SE, discuss the impact of pilot allocation and user density, and study the behaviors as $N \rightarrow \infty$.

\subsection{Pilot-Based Channel Estimation}

BS $j$ can use its multitude of antennas for adaptive receive beamforming, which can amplify useful signals and reject interfering signals. This requires, however, some knowledge of the effective user channels $\sqrt{p_{lk}} \vect{h}_{jlk}$ (for all $l$ and $k$). Such CSI is typically acquired by pilot signaling, where the UEs send known signals. This is a challenging task in networks where the transmission resources are reused across cells, because the pilot signals are then affected by strong inter-cell interference. This so-called \emph{pilot contamination} limits the quality of the acquired CSI and the ability to reject inter-cell interference (unless certain intricate algorithms can be used \cite{Mueller2013a}).

The impact of pilot contamination is usually studied under the assumption that exactly the same pilot signals are used in all cells. In contrast, we now derive the main properties of massive MIMO systems (with power control) for arbitrary pilot allocation. The pilot signals span $B$ channel uses, where $1 \leq B \leq T$, and are sent in the beginning of each coherence block.\footnote{The pilot signals need not be synchronized across cells as assumed herein, but there is little to gain from shifting pilot and information signals between cells; this leads to a mix of deterministic pilots and stochastic information at each channel use, but the average pilot contamination will not change in any substantial way \cite[Remark 5]{Ngo2013a}. The effective interference suppression concepts considered in this paper are also harder to implement in such cases.} Each pilot signal can be represented by a deterministic vector $\vect{v} \in \mathbb{C}^B$ and the fixed per-symbol power implies that all entries have unit magnitude. We assume that all pilot signals originate from a predefined \emph{pilot book}
\begin{equation}
\mathcal{V} = \{ \vect{v}_1,\ldots,\vect{v}_B\} \quad \text{where} \quad \vect{v}_{b_1}^{\Htran} \vect{v}_{b_2} = \begin{cases} B, & b_1 = b_2,\\ 0, & b_1 \neq b_2. \end{cases}
\end{equation}
Hence, the $B$ pilot signals form an orthogonal basis and can, for example, be taken as the columns of a discrete Fourier transform (DFT) matrix \cite{Biguesh2004a}. The signal transmitted by UE $k$ in cell $l$ is partitioned as $\vect{x}_{lk}^{\Htran} = [ \vect{v}_{i_{lk}}^{\Htran} \,\, \check{\vect{x}}_{lk}^{\Htran} ]$, where $i_{lk} \in \{1,\ldots,B\}$ is the index of the pilot signal and $\check{\vect{x}}_{lk}^{\Htran} \in \mathbb{C}^{1 \times (T-B)}$ is the information signal.

Using this notation, the received signal at BS $j$ can be partitioned as $\vect{Y}_j = [\tilde{\vect{Y}}_j \,\, \check{\vect{Y}}_j]$ where $\tilde{\vect{Y}}_j \in \mathbb{C}^{N \times B}$ is the received signal during pilot signaling and  $\check{\vect{Y}}_j \in \mathbb{C}^{N \times (T-B)}$ is received during information transmission. According to \eqref{eq:system-model}, these matrices are given by
\begin{align} \label{eq:system-model-divided1}
\tilde{\vect{Y}}_j &=  \sum_{l \in \mathcal{B}} \sum_{k=1}^{K} \sqrt{p_{lk}} \vect{h}_{jlk} \vect{v}_{i_{lk}}^{\Htran} + \tilde{\vect{N}}_{j}  \\
\check{\vect{Y}}_j &=  \sum_{l \in \mathcal{B}} \sum_{k=1}^{K} \sqrt{p_{lk}} \vect{h}_{jlk} \check{\vect{x}}_{lk}^{\Htran} + \check{\vect{N}}_{j}  \label{eq:system-model-divided2}
\end{align}
where also the noise matrix is partitioned as $\vect{N}_j = [\tilde{\vect{N}}_j \,\, \check{\vect{N}}_j]$.

Based on this system model, the next lemma derives the linear minimum mean-squared error (LMMSE) estimator of the effective power-controlled channels, which are defined as $\vect{h}_{jlk}^{\mathrm{eff}} =\sqrt{p_{lk}} \vect{h}_{jlk}$.

\begin{lemma} \label{lemma:LMMSE-estimation}
The LMMSE estimate at BS $j$ of the effective power-controlled  channel $\vect{h}_{jlk}^{\mathrm{eff}}$, for any $l \in \mathcal{B}$ and $k\in \{1,\ldots,K\}$, is
\begin{equation} \label{eq:LMMSE-estimator}
  \hat{\vect{h}}_{jlk}^{\mathrm{eff}} = \frac{d_j(\vect{z}_{lk}) }{ d_l(\vect{z}_{lk})} \left( \vect{v}_{i_{lk}}^{\Htran} \boldsymbol{\Psi}^{-1}_j \kron \vect{I}_N \right) \vectorize(\tilde{\vect{Y}}_j)
\end{equation}
where $\kron$ denotes the Kronecker product, $\vectorize(\cdot)$ is vectorization, and
\begin{align}
\boldsymbol{\Psi}_j &= \sum_{ \ell \in \mathcal{B}} \sum_{m=1}^{K} \frac{d_j(\vect{z}_{\ell m}) }{ d_{\ell}(\vect{z}_{\ell m})}  \vect{v}_{i_{\ell m}} \vect{v}_{i_{\ell m}}^{\Htran} + \frac{\sigma^2}{\rho}  \vect{I}_B \label{eq:Cerror-def}.
\end{align}
The estimation error covariance matrix $\vect{C}_{jlk} \in \mathbb{C}^{N \times N}$ is given by
\begin{equation} \label{eq:LMMSE-error-cov}
\begin{split}
\vect{C}_{jlk} &= \mathbb{E}\left\{ ( \vect{h}_{jlk}^{\mathrm{eff}} -\hat{\vect{h}}_{jlk}^{\mathrm{eff}} )( \vect{h}_{jlk}^{\mathrm{eff}} - \hat{\vect{h}}_{jlk}^{\mathrm{eff}} )^{\Htran}   \right\} \\ &= \rho \frac{d_j(\vect{z}_{lk}) }{ d_l(\vect{z}_{lk})} \left( 1 - \frac{d_j(\vect{z}_{lk}) }{ d_l(\vect{z}_{lk})} \vect{v}_{i_{lk}}^{\Htran} \boldsymbol{\Psi}^{-1}_j \vect{v}_{i_{lk}} \right) \vect{I}_N
\end{split}
\end{equation}
and the mean-squared error (MSE) is $\mathrm{MSE}_{jlk} = \tr( \vect{C}_{jlk})$.
\end{lemma}
\begin{IEEEproof}
The expression for an LMMSE estimator is $\hat{\vect{h}}^{\mathrm{eff}}_{jlk} = \mathbb{E}\{ \vect{h}^{\mathrm{eff}}_{jlk}  \vectorize(\tilde{\vect{Y}}_j)^{\Htran} \}  \left( \mathbb{E}\{  \vectorize(\tilde{\vect{Y}}_j) \vectorize(\tilde{\vect{Y}}_j)^{\Htran} \}  \right)^{-1}   \vectorize(\tilde{\vect{Y}}_j)$ \cite[Chapter 12]{Kay1993a}, where the expectations are with respect to channel realizations. The lemma follows from direct algebraic computation.
\end{IEEEproof}

There are two important differences between Lemma \ref{lemma:LMMSE-estimation} and the channel estimators that are conventionally used in the massive MIMO literature: 1) we estimate the effective channels including power control; and 2) the estimator supports arbitrary pilot allocation.

Since the pilot signals in $\mathcal{V}$ form an orthogonal basis, we note that
\begin{equation} \label{eq:pilot-sequence-expression}
\vect{v}_{i_{lk}}^{\Htran} \boldsymbol{\Psi}^{-1}_j = \frac{1}{\sum_{\ell \in \mathcal{B}} \sum_{m=1}^{K} \frac{d_j(\vect{z}_{\ell m}) }{ d_{\ell}(\vect{z}_{\ell m})} \vect{v}_{i_{lk}}^{\Htran} \vect{v}_{i_{\ell m}} + \frac{\sigma^2}{\rho}} \vect{v}_{i_{lk}}^{\Htran}
\end{equation}
which in conjunction with the error covariance matrix in \eqref{eq:LMMSE-error-cov} reveals how the estimation error depends on the allocation of pilot signals to the UEs (i.e., which of the products $\vect{v}_{i_{lk}}^{\Htran} \vect{v}_{i_{\ell m}} $ that are non-zero).

Although Lemma \ref{lemma:LMMSE-estimation} allows for estimation of all channel vectors, each BS can only resolve $B$ different vectors since there are only $B$ orthogonal pilot signals. To show this, we define the $N \times B$ matrix
\begin{equation}
\widehat{\vect{H}}_{\mathcal{V},j} = \left[  \left( \vect{v}_1^{\Htran} \boldsymbol{\Psi}^{-1}_j \kron \vect{I}_N \right) \vectorize(\tilde{\vect{Y}}_j), \ldots, \left( \vect{v}_B^{\Htran} \boldsymbol{\Psi}^{-1}_j \kron \vect{I}_N \right) \vectorize(\tilde{\vect{Y}}_j) \right]
\end{equation}
using all $B$ pilot signals. The channel estimate in \eqref{eq:LMMSE-estimator} for UE $k$ in cell $l$ can be written as $\hat{\vect{h}}_{jlk}^{\mathrm{eff}} = \frac{d_j(\vect{z}_{lk}) }{ d_l(\vect{z}_{lk})} \widehat{\vect{H}}_{\mathcal{V},j} \vect{e}_{i_{lk}}$, where $\vect{e}_{i}$ denotes the $i$th column of the $B \times B$ identity matrix $\vect{I}_B$. This is the essence of pilot contamination; BSs cannot tell apart UEs that use the same pilot signal and thus cannot reject the corresponding interference.

\subsection{Achievable Spectral Efficiency by Receive Beamforming}

The channel estimates allow each BS to semi-coherently detect the information signals from its UEs. In particular, we assume that BS $j$ applies a receive beamforming vector $\vect{g}_{jk} \in \mathbb{C}^{N}$ to the received signal (i.e., $\vect{g}_{jk}^{\Htran} \check{\vect{Y}}_j$) to amplify the signal from its $k$th UE and reject inter-user interference. The interference rejection in massive MIMO systems can be either \emph{passive} or \emph{active}.
The canonical example of passive rejection is maximum ratio combining (MRC), defined as
\begin{equation}
\vect{g}_{jk}^{\mathrm{MRC}} = \widehat{\vect{H}}_{\mathcal{V},j} \vect{e}_{i_{jk}} = \hat{\vect{h}}_{jjk}^{\mathrm{eff}},
\end{equation}
which maximizes the gain of the useful signal and relies on that interfering channels are quasi-orthogonal to $\hat{\vect{h}}_{jjk}^{\mathrm{eff}}$ when $N$ is large.

In contrast, active rejection is achieved by making the receive beamforming as orthogonal to the interfering channels as possible. We propose a new \emph{pilot-based zero-forcing combining} (P-ZFC), defined as
\begin{equation}
\vect{g}_{jk}^{\mathrm{P}\text{-}\mathrm{ZFC}} = \widehat{\vect{H}}_{\mathcal{V},j}(\widehat{\vect{H}}_{\mathcal{V},j}^{\Htran} \widehat{\vect{H}}_{\mathcal{V},j} )^{-1} \vect{e}_{i_{jk}}.
\end{equation}
In contrast to classical zero-forcing receivers, which only try to orthogonalize the $K$ intra-cell channels, P-ZFC exploits that all the $B$ estimated channel directions in $\widehat{\vect{H}}_{\mathcal{V},j}$ are known at BS $j$ and orthogonalize all of them to also mitigate parts of the inter-cell interference. The cost is a loss in array gain of $B$ instead of $K$.

MRT and P-ZFC are the receive beamforming schemes considered in this paper, while other schemes are left for future work. The next theorem provides closed-form expressions for the achievable uplink SEs and effective signal-to-interference-and-noise ratios (SINRs).

\begin{figure*}[t!]
\begin{align} \label{eq:achievable-SINR-MRC}
\overline{\mathrm{SINR}}_{jk}^{\mathrm{MRC}} &=  \frac{ B }{ \left( \fracSum{l \in \mathcal{B} }\mu_{jl}^{(1)} \frac{ K }{ N} + \frac{\sigma^2}{  N \rho } \right)\left(  \fracSum{\ell \in \mathcal{B} } \fracSumtwo{m=1}{K} \mu_{jl}^{(1)} \vect{v}_{i_{jk}}^{\Htran}\vect{v}_{i_{\ell m}} + \frac{\sigma^2}{  \rho} \right)
+  \fracSum{l \in \mathcal{B} } \fracSumtwo{m=1}{K} \left( \mu^{(2)}_{jl} + \frac{\mu^{(2)}_{jl}-\left( \mu^{(1)}_{jl} \right)^2}{N} \right)  \vect{v}_{i_{jk}}^{\Htran}\vect{v}_{i_{l m}} -B  } \\ \label{eq:achievable-SINR-ZF}
\overline{\mathrm{SINR}}_{jk}^{\mathrm{P}\text{-}\mathrm{ZFC}} \! &=  \frac{ B }{ \! \fracSum{l \in \mathcal{B}} \fracSumtwo{m=1}{K}
\!\! \left( \! \mu^{(2)}_{jl} \!+\! \frac{\mu^{(2)}_{jl}-\left( \mu^{(1)}_{jl} \right)^2}{N-B} \! \right) \! \vect{v}_{i_{jk}}^{\Htran}\vect{v}_{i_{l m}}
\!+ \!\! \left( \fracSum{l \in \mathcal{B} } \fracSumtwo{m=1}{K}   \mu_{jl}^{(1)} \!\!
\left( \!
1-
\frac{ B \mu_{j l}^{(1)} }{  \fracSum{\ell \in \mathcal{B} } \fracSumtwo{\tilde{m}=1}{K} \mu_{j \ell}^{(1)} \vect{v}_{i_{lm}}^{\Htran}\vect{v}_{i_{\ell \tilde{m}}} + \frac{\sigma^2}{  \rho} }
\! \right) \!\!
+  \frac{\sigma^2}{ \rho } \! \right)\! \!\! \Bigg( \! \frac{ \fracSum{\ell \in \mathcal{B} } \fracSumtwo{m=1}{K} \mu_{jl}^{(1)} \vect{v}_{i_{jk}}^{\Htran}\vect{v}_{i_{\ell m}} + \frac{\sigma^2}{  \rho} }{N-B} \Bigg) \!- \! B }
\end{align} \vskip-2mm
\hrulefill
\vskip-5mm
\end{figure*}

\begin{theorem} \label{theorem-achievable-rate}
An ergodic achievable SE in cell $j$ is \vspace{-1mm}
\begin{equation}
\mathrm{SE}_j = \sum_{k=1}^{K} \left( 1-\frac{B}{T} \right) \log_2(1+ \overline{\mathrm{SINR}}_{jk}) \quad \text{[bit/s/Hz/cell]}
\end{equation} \vskip-1mm
\noindent where the SINR for UE $k$, $\overline{\mathrm{SINR}}_{jk}$, is given in \eqref{eq:achievable-SINR-MRC} for MRC and \eqref{eq:achievable-SINR-ZF} for P-ZFC at the top of the page. The following notation is used:
\begin{align} \label{eq:mu-definition}
\mu^{(\gamma)}_{jl} &= \mathbb{E}_{\vect{z}_{lm}} \left\{ \left( \frac{d_j(\vect{z}_{lm}) }{ d_l(\vect{z}_{lm})} \right)^{\gamma} \right\} \quad \text{for} \,\,\, \gamma=1,2.
\end{align}
\end{theorem}
\begin{IEEEproof}
The SEs follow from treating interference and channel uncertainty as worst-case Gaussian noise (cf.~\cite{Marzetta2010a,Jose2011b,Hoydis2013a}) and taking variations in interference power (due to random UE positions) into account in the coding. Details are omitted due to limited space.
\end{IEEEproof}

The closed-form SE expressions in Theorem \ref{theorem-achievable-rate} are only functions of the pilot allocation and the propagation parameters $\mu^{(1)}_{jl},\mu^{(2)}_{jl}$. The latter are the average ratio between the channel variance to BS $j$ and the channel variance to BS $l$, for an arbitrary UE in cell $l$, and the average second-order moment of this ratio, respectively. The SE expressions manifest the importance of pilot allocation, since all interference terms contain inner products of pilot signals that are either zero or $B$. Since it is hard to gain insights directly from the SINR expressions in \eqref{eq:achievable-SINR-MRC} and \eqref{eq:achievable-SINR-ZF}, we look at the asymptotic limit.

\begin{corollary}
When $N \rightarrow \infty$ (with $K,B \leq T < \infty$), the SEs with MRT and P-ZFC approach the same limit:
\begin{equation} \label{eq:asymptotic-SINR}
\overline{\mathrm{SINR}}_{jk}^{\mathrm{MRC}}, \overline{\mathrm{SINR}}_{jk}^{\mathrm{P}\text{-}\mathrm{ZFC}} \rightarrow \frac{ B }{ \fracSum{l \in \mathcal{B} } \fracSumtwo{m=1}{K} \mu^{(2)}_{jl}  \vect{v}_{i_{jk}}^{\Htran}\vect{v}_{i_{l m}} -B  }.
\end{equation}
\end{corollary}

In order to maximize the asymptotic SINR in \eqref{eq:asymptotic-SINR}, we should allocate the pilot signals such that $\vect{v}_{i_{jk}}^{\Htran}\vect{v}_{i_{\ell m}}=0$ whenever $\mu^{(2)}_{jl}$ is large.
If $\beta = \frac{B}{K}$ is an integer, this amounts to allocating orthogonal pilots among the UEs in each cell and making sure that only $\frac{1}{\beta}$ of the interfering cells reuse these pilots. We refer to this as \emph{fractional pilot reuse}.
An explicit example is provided in the next section for hexagonal cells. The following result holds for any cell structure.

\begin{corollary} \label{cor:SE-maximization}
Let $B = K \beta$ for some integer $\beta$, assume orthogonal intra-cell pilot signals, and let $\mathcal{B}_j \subset \mathcal{B}$ be the set of cells that use the same pilots as cell $j$.
The asymptotic SE for cell $j$ becomes
\begin{equation} \label{eq:asymptotic-SE}
\mathrm{SE}_j^{\infty} = K \left( 1-\frac{K \beta}{T} \right) \log_2 \bigg(1+ \frac{ 1 }{ \sum_{l \in \mathcal{B}_j \setminus \{ j\} } \mu^{(2)}_{jl} } \bigg),
\end{equation}
when $N \rightarrow \infty$. This SE is maximized jointly for all cells by $K^\star = \left\lfloor \frac{T}{2 \beta} \right\rfloor$ or $K^\star = \left\lceil \frac{T}{2 \beta} \right\rceil$ scheduled users (one of the closest integers).
\end{corollary}
\begin{IEEEproof}
The logarithmic part of \eqref{eq:asymptotic-SE} is independent of $K$, while the concave pre-log factor $K \left( 1\!-\!\frac{K \beta}{T} \right)$ is maximized by $K \!=\! \frac{T}{2 \beta}$.
\end{IEEEproof}

Corollary \ref{cor:SE-maximization} is a main contribution of this paper and proves that the number of scheduled UEs should be proportional to the coherence block length $T$ (when $N$ is large enough); for example, we get $K^\star=\frac{T}{2}$ for $\beta = 1$ and $K^\star=\frac{T}{6}$ for $\beta = 3$. Since both $T=500$ and $T = 10000$ are reasonable coherence block lengths in practice, depending on the UE mobility, this means that we should schedule hundreds or even thousands of simultaneous UEs in order to be optimal. This is only possible if the UE selection policy is simple and scalable. If $K^\star = \frac{T}{2 \beta}$ is an integer, the asymptotically optimal SE is \vspace{-2mm}
\begin{equation} \label{eq:asymptotic-SE-optimized}
\mathrm{SE}_j^{\infty} = \frac{T}{4 \beta} \log_2 \bigg(1+ \frac{ 1 }{ \sum_{l \in \mathcal{B}_j \setminus \{ j\} } \mu^{(2)}_{jl} } \bigg)
\end{equation} \vskip -1mm
\noindent and increases linearly with $T$ (in the large-$N$ regime).

Interestingly, the optimal scheduling gives $B = \frac{T}{2}$ for any $\beta$, which means that \emph{half} the coherence block is used for pilot transmission. This is an extraordinary fact that bears some similarity  with results in \cite{Zheng2002a,Hassibi2003a} for block-fading noncoherent point-to-point MIMO channels, where the maximal degrees of freedom (DoF) are $\frac{T}{4}$ and are achieved by having $\frac{T}{2}$ transmit/receive antennas and using pilots of the same length. The fundamental difference is that the DoF concept, where unbounded SE is achieved at high SNRs, does not apply to cellular networks \cite{Lozano2013a}. Instead, the pre-log factor $\frac{T}{4 \beta}$ in \eqref{eq:asymptotic-SE-optimized} may be interpreted as the relative improvement in SE that can be achieved by aggressive UE scheduling in massive MIMO systems.

We have now established the asymptotically optimal number of scheduled UEs, as $N \rightarrow \infty$. To investigate the impact on practical systems with finite $N$, the next section considers a certain network.

\section{Optimizing Number of Users in Hexagonal Networks}

The concept of cellular communications has been around for decades \cite{Macdonald1979a}. Although practical deployments have irregular cells, it is common practice to establish general properties by analyzing symmetric networks where the cells are regular hexagons \cite{Cox1982a}.

\begin{figure}[!hb]
\begin{center} \vskip-2mm
\includegraphics[width=0.96\columnwidth]{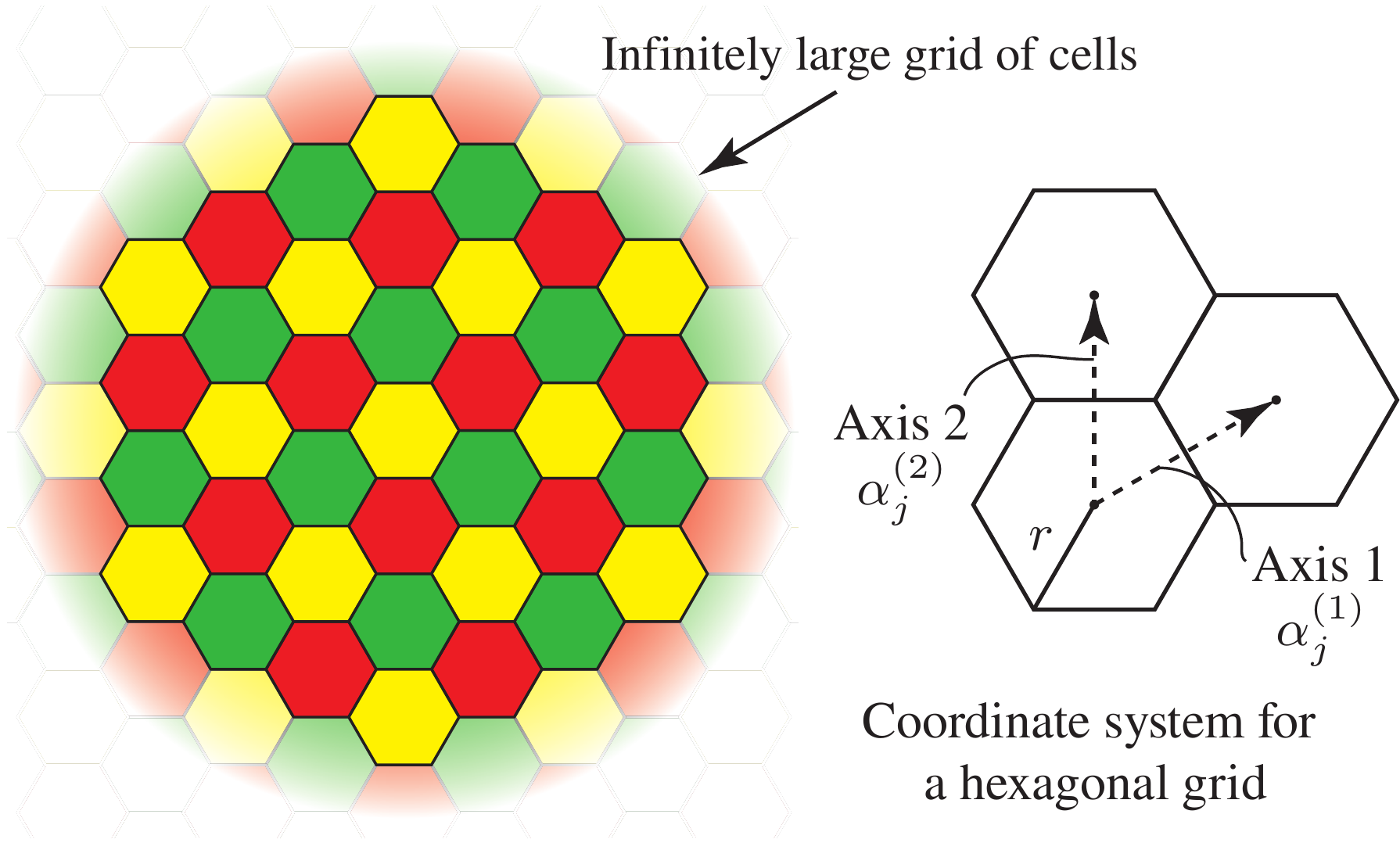}
\end{center}\vskip-5mm
\caption{Part of a hexagonal network, colored for pilot reuse $\beta = 3$.} \label{figure_hexagonal} \vskip-2mm
\end{figure}

In this section, we consider the symmetric network depicted in Fig.~\ref{figure_hexagonal} with hexagonal cells of radius $r>0$ (the distance from center to corners). The hexagonal grid is infinitely large, to avoid edge effect and giving all cells the same properties. Each cell can be uniquely indexed by a pair of integers $\alpha_j^{(1)},\alpha_j^{(2)} \in \mathbb{Z}$, where $\mathbb{Z}$ is the set of all integers. This integer pair specifies the location of BS $j$ \cite{Macdonald1979a}:
\begin{equation}
\vect{b}_j =
\begin{bmatrix}
3r/2 \\
\sqrt{3} r/2
\end{bmatrix} \alpha_j^{(1)}
+\begin{bmatrix}
0 \\
\sqrt{3}r
\end{bmatrix} \alpha_j^{(2)}  \in \mathbb{R}^2.
\end{equation}
We assume a classic pathloss model where the variance of the channel attenuation in \eqref{eq:channel-distribution} is $d_j(\vect{z}) = \frac{C}{\| \vect{z} - \vect{b}_j\|^{\kappa}}$, where $\|\cdot\|$ is the Euclidean norm, $C>0$ is a reference value, and $\kappa\geq 2$ is the pathloss exponent. These assumptions allow us to compute $\mu^{(\gamma)}_{jl}$ in \eqref{eq:mu-definition} as
\begin{equation} \label{eq:mu-computation}
\mu^{(\gamma)}_{jl}  = \mathbb{E}_{\vect{z}_{lm}} \! \left\{ \left( \frac{d_j(\vect{z}_{lm}) }{ d_l(\vect{z}_{lm})} \right)^{\gamma} \right\} = \mathbb{E}_{\vect{z}_{lm}} \! \left\{ \left(\frac{\| \vect{z}_{lm} \!-\! \vect{b}_l\|}{\| \vect{z}_{lm} \!-\! \vect{b}_j\|} \right)^{\kappa \gamma} \right\}
\end{equation}
for any UE distributions in the cells. We notice that $C$ and $r$ cancel out in \eqref{eq:mu-computation}, if the UE distribution is invariant to the radius. Since the power control makes the SINRs in Theorem \ref{theorem-achievable-rate} independent of the UE's position, we only need to define the parameter $\rho$; the average SNR (over fading) between any UE and any antenna at its serving BS. This parameter is set to 10 dB in the simulations, which is not particularly large but allows for decent channel estimation accuracy and convergence speed to the large-$N$ regime; see \cite{Bjornson2014a}.

\begin{figure}[t!]
\begin{subfigure}[b]{\columnwidth}
\includegraphics[width=\columnwidth]{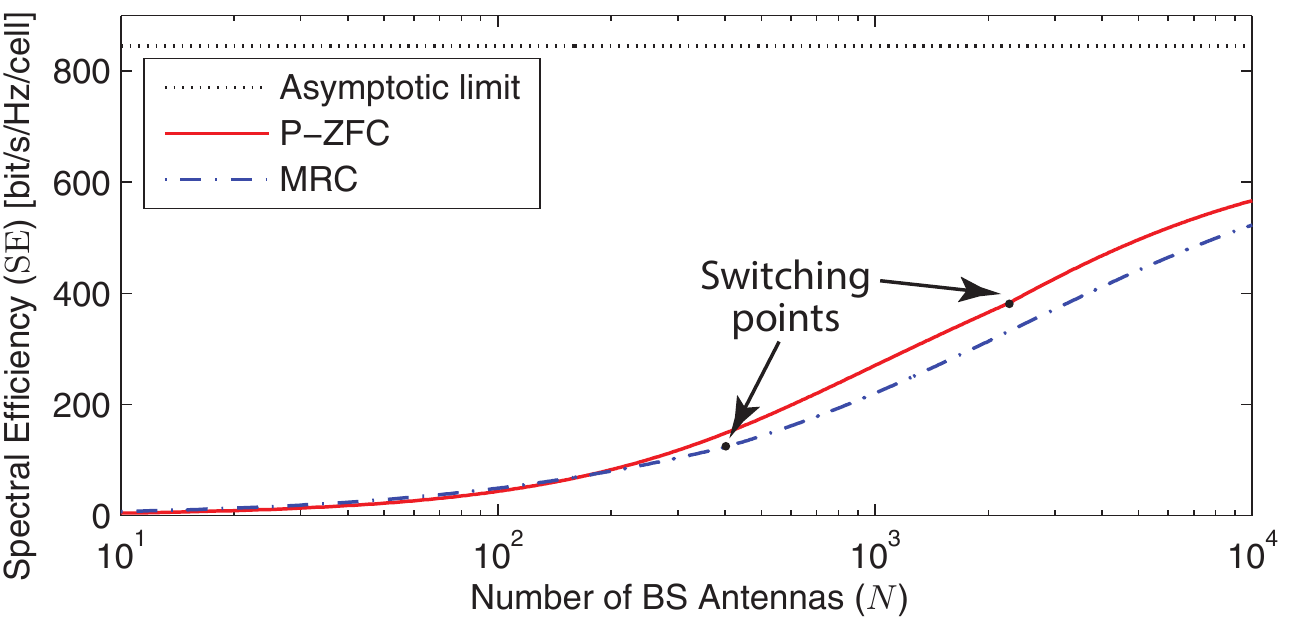} \vspace{-4mm}
\caption{Optimized SE per cell.} \label{figure_SE_mean}
\end{subfigure}
\begin{subfigure}[b]{\columnwidth}
\includegraphics[width=\columnwidth]{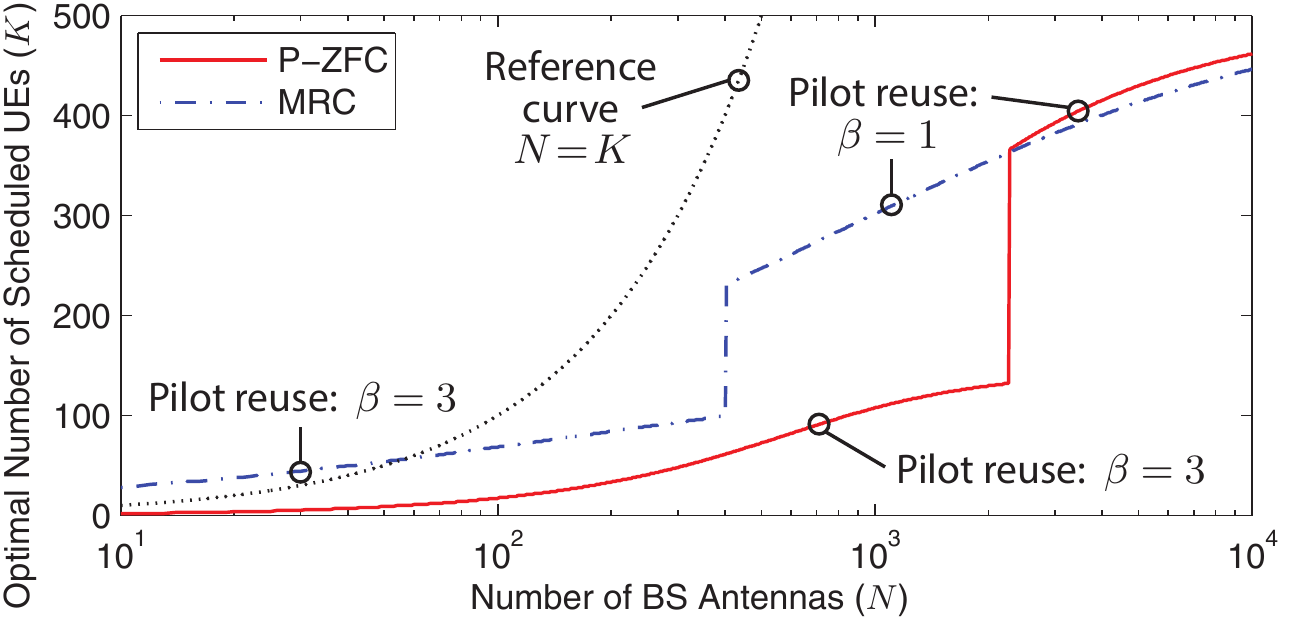} \vspace{-4mm}
\caption{Optimal number of scheduled UEs, $K^\star$, per cell.} \label{figure_K_mean}
\end{subfigure}
\caption{Simulation results with average inter-cell interference.} \label{figure_simulation_mean} \vspace{-5mm}
 \end{figure}

Although all cells reuse all time/frequency resources, we consider pilot books of size $B = \beta K$ to allow for fractional pilot reuse that avoids pilot contamination from neighboring cells. The hexagonal grid only allows for some reuse factors: $\beta \in  \{1,3,4,7, \ldots \}$ \cite{Macdonald1979a}.

\subsection{Simulation Results}

We simulate the uplink SE in an arbitrary cell on the hexagonal grid in Fig.~\ref{figure_hexagonal} and take all non-negligible interference into account. For each number of antennas, $N$, we optimize the SE with respect to the number of UEs $K$ and pilot reuse factor $\beta$ (which determine $B = \beta K$). We set the coherence block length to $T = 1000$ (e.g., achieved by  10 ms coherence time and 100 kHz coherence bandwidth), use the whole block for uplink, and pick $\kappa = 3.5$ as pathloss exponent.

\begin{figure}[t!]
\begin{subfigure}[b]{\columnwidth}
\includegraphics[width=\columnwidth]{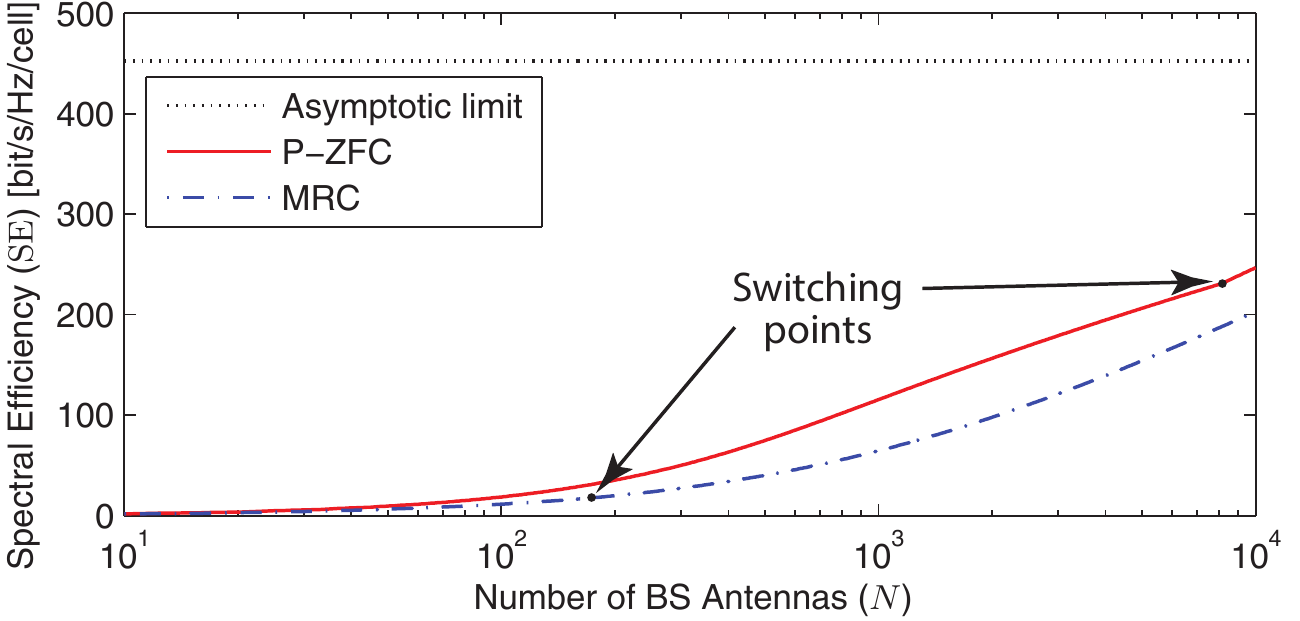} \vspace{-4mm}
\caption{Optimized SE per cell.} \label{figure_SE_worst}
\end{subfigure}
\begin{subfigure}[b]{\columnwidth}
\includegraphics[width=\columnwidth]{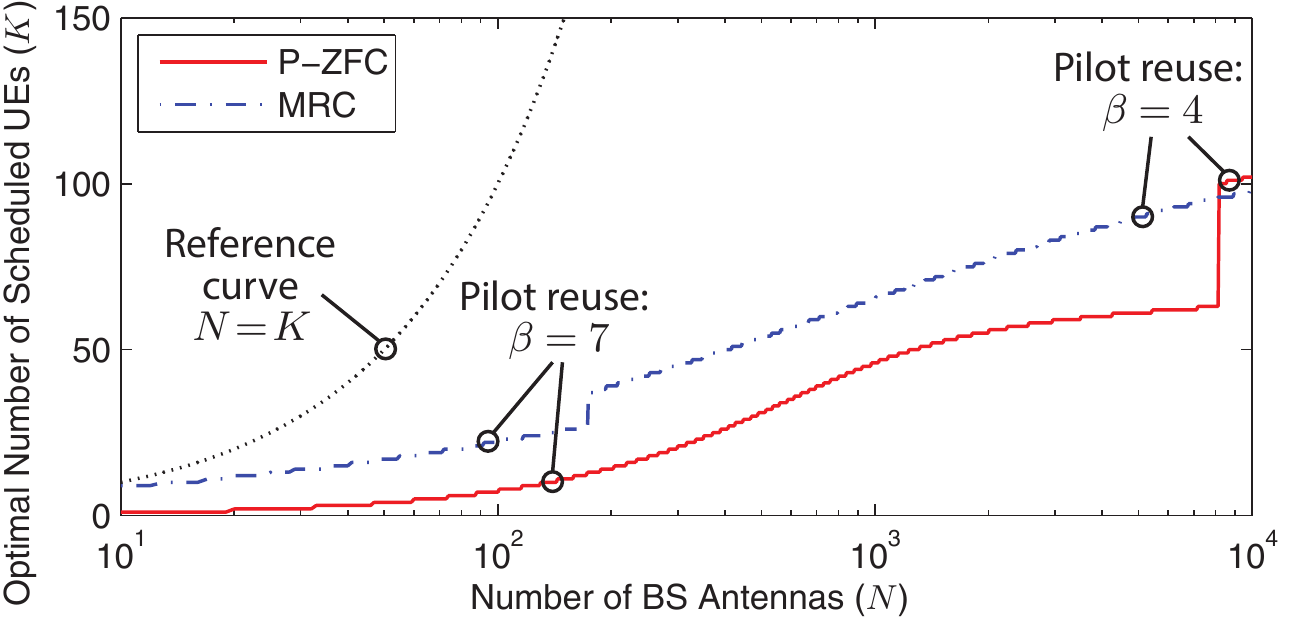} \vspace{-4mm}
\caption{Optimal number of scheduled UEs, $K^\star$, per cell.} \label{figure_K_worst}
\end{subfigure}
\caption{Simulation results with worst-case inter-cell interference.} \label{figure_simulation_worst} \vspace{-5mm}
 \end{figure}

We compute $\mu^{(1)}_{jl}$ and $\mu^{(2)}_{jl}$ in two different ways: 1) by averaging over uniform UE locations in all cells (restricting all UEs to be at least $0.14r$ from its BS); and 2) worst-case scenario where all UEs in other cells are at the cell edge closest to BS $j$ (for each $j$). The former is optimistic since the variations in interference power from the average number can be large, while the latter is overly pessimistic since the UEs cannot all be at the worst locations simultaneously. The average case is shown in Fig.~\ref{figure_simulation_mean} and the worst case in Fig.~\ref{figure_simulation_worst}.

Interestingly, the optimized SEs are similar for MRC and P-ZFC for $10 \leq N \leq 200$ antennas, but are achieved by scheduling very different numbers of UEs. MRC gives low per-user SEs to many UEs (sometimes more than $N$), while P-ZFC gives higher per-user SEs to fewer UEs. The latter is made possible by the active interference rejection in P-ZFC. For larger $N$, P-ZFC has a distinct advantage in terms of per-cell SE; particularly for worst case interference where there is a greater need for active interference rejection. Both schemes use fractional pilot reuse at small $N$ to reduce the inter-cell interference, while the optimal $\beta$ reduces at large $N$ since the UEs' channels are asymptotically quasi-orthogonal. Higher pilot reuse factors $\beta$ are needed in the worst case, than in the average case. None of the schemes comes close to the asymptotic limit in \eqref{eq:asymptotic-SE-optimized} at $N=10^4$ antennas; one typically needs $N=10^6$ to reach the limit.

The number of scheduled UEs is very different between the mean and worst case interference. In Fig.~\ref{figure_simulation_mean}, where the inter-cell interference is relatively mild, it can be preferable to schedule hundreds of UEs; both schemes approach the maximum of $K = \frac{T}{2} = 500$ as $N$ grows. For the worst-case interference in Fig.~\ref{figure_simulation_worst}, it is preferable to schedule fewer UEs, but there can still be many tens of active UEs per cell.

\vspace{-1mm}

\section{Conclusion}

\enlargethispage{1mm}

This paper investigated how many UEs, $K$, that should be scheduled in massive MIMO (with power control) to maximize the SE per cell for a fixed $N$. The optimal $K$ approaches $\frac{T}{2\beta}$ as $N \rightarrow \infty$, but can be different at finite $N$. Different schemes prefer different $K$ and $\beta$, thus it is hard to make fair comparisons. The common rule-of-thumb of $N \gg K$ holds true when $N \gg \frac{T}{2}$, but not when the average inter-cell interference is weak. By simulations, a $100 \times$ increase in SE over the IMT-Advanced  requirement of 2.25 bit/s/Hz/cell was demonstrated. Taking the ratio between SE and $K$ reveals that this is achieved by having an SE per UE in the conventional range (1-3 bit/s/Hz/user), but schedule unconventionally many UEs per cell.

\newpage

\enlargethispage{-10.5cm}

\bibliographystyle{IEEEtran}
\bibliography{IEEEabrv,refs}

\begin{thebibliography}{10}
\providecommand{\url}[1]{#1}
\csname url@samestyle\endcsname
\providecommand{\newblock}{\relax}
\providecommand{\bibinfo}[2]{#2}
\providecommand{\BIBentrySTDinterwordspacing}{\spaceskip=0pt\relax}
\providecommand{\BIBentryALTinterwordstretchfactor}{4}
\providecommand{\BIBentryALTinterwordspacing}{\spaceskip=\fontdimen2\font plus
\BIBentryALTinterwordstretchfactor\fontdimen3\font minus
  \fontdimen4\font\relax}
\providecommand{\BIBforeignlanguage}[2]{{%
\expandafter\ifx\csname l@#1\endcsname\relax
\typeout{** WARNING: IEEEtran.bst: No hyphenation pattern has been}%
\typeout{** loaded for the language `#1'. Using the pattern for}%
\typeout{** the default language instead.}%
\else
\language=\csname l@#1\endcsname
\fi
#2}}
\providecommand{\BIBdecl}{\relax}
\BIBdecl

\bibitem{Nokia2011}
{Nokia Siemens Networks}, ``2020: Beyond {4G} radio evolution for the {Gigabit}
  experience,'' White Paper, Tech. Rep., 2011.

\bibitem{Marzetta2010a}
T.~L. Marzetta, ``Noncooperative cellular wireless with unlimited numbers of
  base station antennas,'' \emph{{IEEE} Trans. Wireless Commun.}, vol.~9,
  no.~11, pp. 3590--3600, Nov. 2010.

\bibitem{Baldemair2013a}
R.~Baldemair, E.~Dahlman, G.~Fodor, G.~Mildh, S.~Parkvall, Y.~Selen,
  H.~Tullberg, and K.~Balachandran, ``Evolving wireless communications:
  Addressing the challenges and expectations of the future,'' \emph{{IEEE} Veh.
  Technol. Mag.}, vol.~8, no.~1, pp. 24--30, Mar. 2013.

\bibitem{Boccardi2014a}
F.~Boccardi, R.~Heath, A.~Lozano, T.~Marzetta, and P.~Popovski, ``Five
  disruptive technology directions for {5G},'' \emph{{IEEE} Commun. Mag.},
  vol.~52, no.~2, pp. 74--80, Feb. 2014.

\bibitem{Larsson2014a}
E.~G. Larsson, F.~Tufvesson, O.~Edfors, and T.~L. Marzetta, ``Massive {MIMO}
  for next generation wireless systems,'' \emph{{IEEE} Commun. Mag.}, vol.~52,
  no.~2, pp. 186--195, Feb. 2014.

\bibitem{Jose2011b}
J.~Jose, A.~Ashikhmin, T.~L. Marzetta, and S.~Vishwanath, ``Pilot contamination
  and precoding in multi-cell {TDD} systems,'' \emph{{IEEE} Trans. Commun.},
  vol.~10, no.~8, pp. 2640--2651, Aug. 2011.

\bibitem{Hoydis2013a}
J.~Hoydis, S.~ten Brink, and M.~Debbah, ``Massive {MIMO} in the {UL/DL} of
  cellular networks: How many antennas do we need?'' \emph{{IEEE} J. Sel. Areas
  Commun.}, vol.~31, no.~2, pp. 160--171, Feb. 2013.

\bibitem{Ngo2013a}
H.~Q. Ngo, E.~G. Larsson, and T.~L. Marzetta, ``Energy and spectral efficiency
  of very large multiuser {MIMO} systems,'' \emph{{IEEE} Trans. Commun.},
  vol.~61, no.~4, pp. 1436--1449, Apr. 2013.

\bibitem{Bjornson2015a}
E.~Bj{\"{o}}rnson, L.~Sanguinetti, J.~Hoydis, and M.~Debbah, ``Optimal design
  of energy-efficient multi-user {MIMO} systems: Is massive {MIMO} the
  answer?'' \emph{{IEEE} Trans. Wireless Commun.}, submitted, Available:
  http://arxiv.org/abs/1403.6150.

\bibitem{Bjornson2014a}
E.~Bj{\"{o}}rnson, J.~Hoydis, M.~Kountouris, and M.~Debbah, ``Massive {MIMO}
  systems with non-ideal hardware: Energy efficiency, estimation, and capacity
  limits,'' \emph{{IEEE} Trans. Inf. Theory}, to appear.

\bibitem{Bjornson2015b}
E.~Bj{\"{o}}rnson, M.~Matthaiou, and M.~Debbah, ``Massive {MIMO} with arbitrary
  non-ideal arrays: Hardware scaling laws and circuit-aware design,''
  \emph{{IEEE} Trans. Wireless Commun.}, submitted, Available:
  http://arxiv.org/abs/1409.0875.

\bibitem{Huh2012a}
H.~Huh, G.~Caire, H.~Papadopoulos, and S.~Ramprashad, ``Achieving ``massive
  {MIMO}'' spectral efficiency with a not-so-large number of antennas,''
  \emph{{IEEE} Trans. Wireless Commun.}, vol.~11, no.~9, pp. 3226--3239, Sept.
  2012.

\bibitem{Li2012a}
M.~Li, Y.-H. Nam, B.~Ng, and J.~Zhang, ``A non-asymptotic throughput for
  massive {MIMO} cellular uplink with pilot reuse,'' in \emph{Proc.~IEEE
  Globecom}, 2012.

\bibitem{Mueller2013a}
R.~M\"{u}ller, M.~Vehkaper\"{a}, and L.~Cottatellucci, ``Blind pilot
  decontamination,'' in \emph{Proc.~ITG Workshop on Smart Antennas (WSA)},
  2013.

\bibitem{Yin2013a}
H.~Yin, D.~Gesbert, M.~Filippou, and Y.~Liu, ``A coordinated approach to
  channel estimation in large-scale multiple-antenna systems,'' \emph{{IEEE} J.
  Sel. Areas Commun.}, vol.~31, no.~2, pp. 264--273, Feb. 2013.

\bibitem{Li2013a}
M.~Li, S.~Jin, and X.~Gao, ``Spatial orthogonality-based pilot reuse for
  multi-cell massive {MIMO} transmission,'' in \emph{Proc.~WCSP}, 2013.

\bibitem{Gao2015a}
\BIBentryALTinterwordspacing
X.~Gao, O.~Edfors, F.~Rusek, and F.~Tufvesson, ``Massive {MIMO} in real
  propagation environments,'' \emph{{IEEE} Trans. Wireless Commun.}, 2014,
  submitted. [Online]. Available: \url{http://arxiv.org/abs/1403.3376}
\BIBentrySTDinterwordspacing

\bibitem{Guo2014a}
K.~Guo, Y.~Guo, G.~Fodor, and G.~Ascheid, ``Uplink power control with {MMSE}
  receiver in multi-cell {MU}-massive-{MIMO} systems,'' in \emph{Proc.~IEEE
  ICC}, 2014.

\bibitem{Biguesh2004a}
M.~Biguesh and A.~B. Gershman, ``Downlink channel estimation in cellular
  systems with antenna arrays at base stations using channel probing with
  feedback,'' \emph{{EURASIP} J. Appl. Signal Process.}, vol. 2004, no.~9, pp.
  1330--1339, 2004.

\bibitem{Kay1993a}
S.~M. Kay, \emph{Fundamentals of Statistical Signal Processing: Estimation
  Theory}.\hskip 1em plus 0.5em minus 0.4em\relax Prentice Hall, 1993.

\bibitem{Zheng2002a}
L.~Zheng and D.~Tse, ``Communication on the {Grassmann} manifold: A geometric
  approach to the noncoherent multiple-antenna channel,'' \emph{{IEEE} Trans.
  Inf. Theory}, vol.~48, no.~2, pp. 359--383, Feb. 2002.

\bibitem{Hassibi2003a}
B.~Hassibi and B.~M. Hochwald, ``How much training is needed in
  multiple-antenna wireless links?'' \emph{{IEEE} Trans. Inf. Theory}, vol.~49,
  no.~4, pp. 951--963, Apr. 2003.

\bibitem{Lozano2013a}
A.~Lozano, R.~Heath, and J.~Andrews, ``Fundamental limits of cooperation,''
  \emph{{IEEE} Trans. Inf. Theory}, vol.~59, no.~9, pp. 5213--5226, Sept. 2013.

\bibitem{Macdonald1979a}
V.~M. Donald, ``The cellular concept,'' \emph{Bell System Technical Journal},
  vol.~58, no. 15-41, pp. 113--381, 1979.

\bibitem{Cox1982a}
D.~Cox, ``Cochannel interference considerations in frequency reuse
  small-coverage-area radio systems,'' \emph{{IEEE} Trans. Commun.}, vol.~30,
  no.~1, pp. 135--142, Jan. 1982.

\end{thebibliography}

\end{document}